\documentclass[slac_one]{revtex4}
\usepackage{graphicx}
\usepackage{fancyhdr}
\usepackage{amsbsy}
\usepackage{amsmath}
\usepackage{amsfonts}
\usepackage{amssymb}
\newcommand{\gev}{\,\mbox{GeV}}
\newcommand{\mev}{\,\mbox{MeV}}
\newcommand{\BDTN}{B \to D \tau \nu}

\newcommand{\BDLN}{B \to D \ell \nu}

\begin{document}

\title{Charged-Higgs effects in \boldmath $B\rightarrow(D)\tau\nu$ \unboldmath decays}
\author{St\'{e}phanie Trine}
\affiliation{Institut f\"{u}r Theoretische Teilchenphysik, Universit\"{a}t Karlsruhe,
D-76128 Karlsruhe, Germany}

\begin{abstract}
We update and compare the capabilities of the purely leptonic mode $B\to\tau\nu$
and the semileptonic mode $B\to D\tau\nu$ in the search for a charged Higgs boson.
\end{abstract}

\maketitle
\thispagestyle{fancy}

\section{INTRODUCTION}

Supersymmetric extensions of the standard model (SM) -- or more generally
extensions that require the existence of at least one additional Higgs doublet
-- generate new flavour-changing interactions already at tree-level via the
exchange of a charged Higgs boson. The coupling of $H^{+}$ to fermions grows with
the fermion mass. It is thus natural to look at (semi)leptonic $B$ decays
with a $\tau$ in the final state to try to uncover this type of effects.
In a two-Higgs-doublet-model (2HDM) of type II, where up-type quarks
get their mass from one of the two Higgs doublets and down-type quarks from the other one,
$H^{+}$ effects are entirely parametrized by
the $H^{+}$ mass, $M_H$, and the ratio of the two Higgs vacuum expectation values,
$\tan\beta=v_{u}/v_{d}$. They can compete with the exchange of a $W^+$  boson
for large values of $\tan\beta$ \cite{Hou93}.
In the minimal supersymmetric extension of the SM (MSSM),
the tree-level type-II structure is spoilt by radiative corrections
involving supersymmetry-breaking terms.
The effective scalar coupling $g_{S}$ then exhibits an additional dependence
on sparticle mass parameters when $\tan\beta$ is large $(q=u,c)$ \cite{AkeroydR03,Pheno}:
\begin{equation}
H_{eff}^{H^{+}}=-2\sqrt{2}G_{F}V_{qb}\frac{m_{b}m_{\tau}}{M_{B}
^{2}}g_{S}\left[  \overline{q}_{L}b_{R}\right]  \left[  \overline{\tau}_{R}
\nu_{L}\right]  +h.c.,\qquad g_{S}=\frac{M_{B}^{2}\tan^{2}\beta}{M_{H}^{2}
}\frac{1}{(1+\varepsilon_{0}\tan\beta)(1+\varepsilon_{\tau}\tan\beta)},
\label{Eq1}
\end{equation}
where $\varepsilon_{0,\tau}$ denote sparticle loop factors.
The correction induced can be of order one. However, the access to the Higgs sector
remains exceptionally clean.
In Eq.(\ref{Eq1}), $g_{S}$ has been normalized such that it gives the fraction of effects
in the $B\rightarrow\tau\nu$ amplitude, which is very sensitive to $H^{+}$ exchange:
$\mathcal{B}(B\rightarrow\tau\nu)/\mathcal{B}(B\rightarrow\tau\nu)^{SM}=|1-g_{S}|^{2}$.
The $B\rightarrow D\tau\nu$ channel is less
sensitive (though better in this respect than other modes such as $B\to D^*\tau\nu$) but,
as we will see, exhibits a number of features
that make it, too, play an important part in the hunt for the charged Higgs~boson.

\section{\boldmath$\mathcal{B}(B\rightarrow D\tau\nu)$
VERSUS $\mathcal{B}(B\rightarrow\tau\nu)$\unboldmath}

\begin{figure}[t]
\centering
\begin{tabular}{cc}
\includegraphics[width=0.43\linewidth,keepaspectratio=true,angle=0]{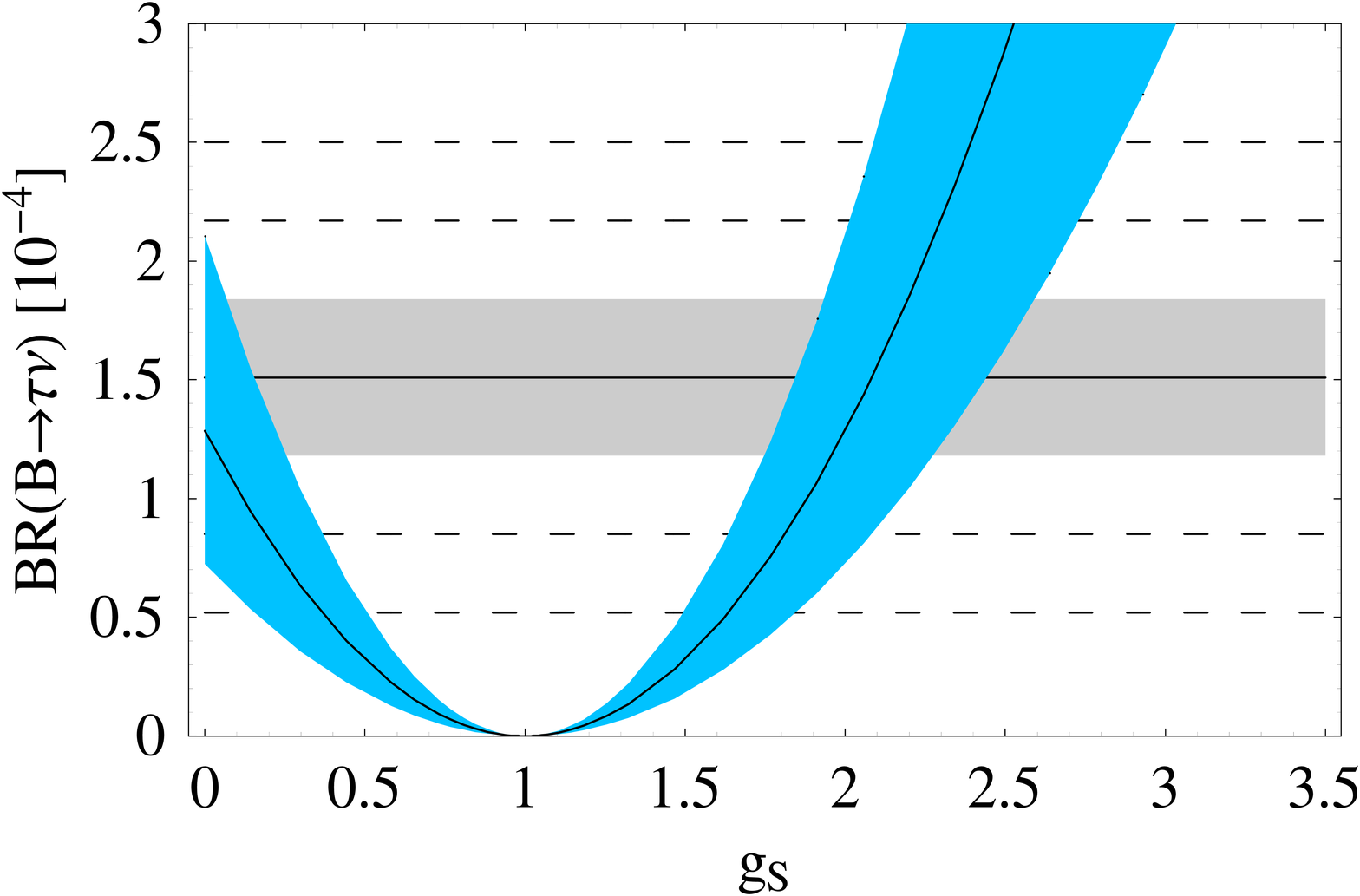} &
\includegraphics[width=0.43\linewidth,keepaspectratio=true,angle=0]{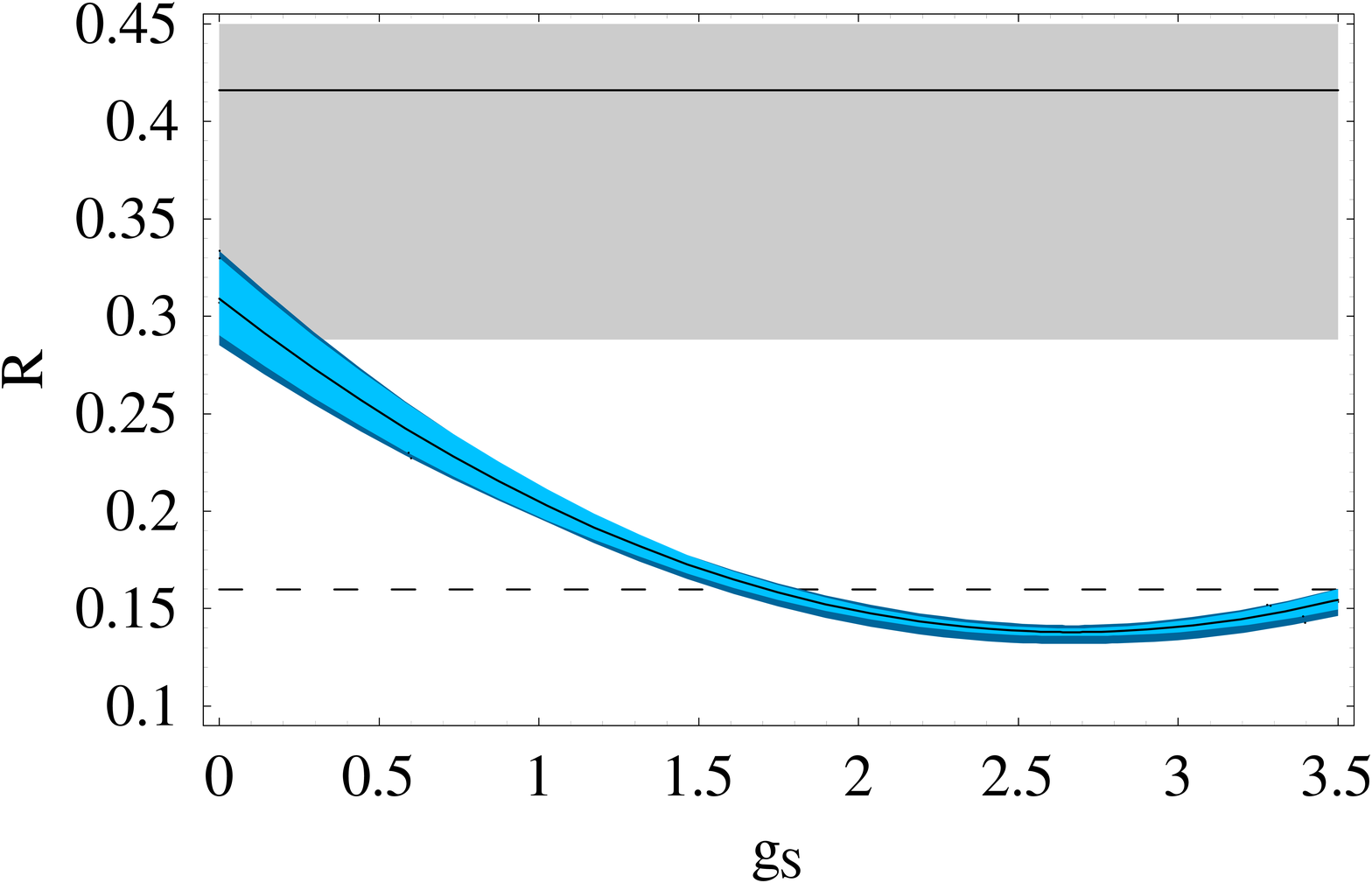} 
\end{tabular}
\caption{Left: $\mathcal{B}(B\rightarrow\tau\nu)$ as a function of $g_{S}$.
Light gray band: $\mathcal{B}^\textrm{exp}=(1.51\pm0.33)\times10^{-4}$ \cite{HFAG08}.
Gray (blue) band: $\mathcal{B}^\textrm{th}$
for $|V_{ub}|=(3.95\pm0.35)\times10^{-3}$ \cite{PDG08}
and $f_B=216\pm38\mev$ \cite{HPQCD05}
(in this last case, we add errors linearly to stay on the conservative side).
Right: $R\equiv\mathcal{B}(\BDTN)/\mathcal{B}(\BDLN)$ as a function of $g_S$.
Light gray band: $R^\textrm{exp}=(41.6\pm11.7\pm5.2)\%$ \cite{BABAR08_BDTauNu}.
Dark gray (dark blue) band: $R^\textrm{th}$ with the HFAG vector form factor before ICHEP08 \cite{HFAG07}.
Gray (blue) band: $R^\textrm{th}$ with the HFAG vector form factor after ICHEP08 \cite{HFAG08}.
The dashed lines indicate the $2$ and $3$-sigma limits.
The ratio $m_c/m_b$ in the $\overline{\textrm{MS}}$ scheme has recently been determined with very high accuracy:
$m_c/m_b=0.2211\pm0.0044$ \cite{mcmb}.
We inflated the error on this number and set $m_c/m_b=0.22\pm0.01$
to reduce the discrepancy with the HFAG estimation \cite{HFAG08}.}
\label{Fig1}
\end{figure}

The current capabilities of $\mathcal{B}(B\rightarrow D\tau\nu)$ and $\mathcal{B}(B\rightarrow\tau\nu)$
to constrain $H^{+}$ effects are compared in Fig.\ref{Fig1} for $g_S\geq0$ (as is typically the case in the MSSM or the 2HDM-II).
The lower sensitivity of the $B\rightarrow D\tau\nu$ mode comes from the different momentum dependence
of the Higgs contribution with respect to the longitudinal $W^+$ one\footnote{Note that the latter,
though helicity-suppressed, is still (slightly) larger than the transverse $W^+$ contribution for all $q^2$ values.}:
$(d\Gamma(B\to D\tau\nu)/dq^2)^{W_{\parallel}^{+}+H^{+}} \propto |1-g_S (q^2/M_B^2)/(1-m_c/m_b)|^2$
with $q\equiv p_B-p_D$.
On the other hand, the theory prediction for $\mathcal{B}(B\rightarrow\tau\nu)$ suffers from large parametric uncertainties
from the CKM matrix element $V_{ub}$ and the $B$ decay constant $f_B$.
In contrast, $V_{cb}$ is known with better than $2\%$ accuracy from inclusive $B\to X_c\ell\nu\ (\ell=e,\mu)$ decays,
$|V_{cb}|=(41.6\pm0.6)\times10^{-3}$ \cite{PDG08},
and the form factors $f_+(q^2)$ and $f_0(q^2)$ describing the $B \to D$ transition are very well under control, as we now discuss in more detail.

To this end, we introduce the following conformal transformation:
\begin{equation}
q^{2}\to z(q^{2},t_{0})\equiv
\frac{\sqrt{(M_{B}+M_{D})^{2}-q^{2}}-\sqrt{(M_{B}+M_{D})^{2}-t_{0}}}{\sqrt{(M_{B}+M_{D})^{2}-q^{2}}+\sqrt{(M_{B}+M_{D})^{2}-t_{0}}},
\label{Eq2}
\end{equation}
which maps the complex $q^2$ plane, cut along $q^2\geq(M_B+M_D)^2$,
onto the disk $|z|<1$. The form factors $f_+$ and $f_0$ are analytic in $z$ in this domain, up to a few subthreshold poles,
and can thus be written as a power series in $z$ after these poles are factored out ($i=+,0$) \cite{zpara}:
\begin{equation}
f_{i}(q^{2})=\frac{1}{P_{i}(q^{2})\phi_{i}(q^{2},t_{0})}\left[  a_{0}^{i}(t_{0})+a_{1}^{i}(t_{0})z(q^{2},t_{0})+...\right],
\label{Eq3}
\end{equation}
where the function $P_{i}$ gathers the pole singularities and an arbitrary analytic function $\phi_{i}$ can be factored out as well.

\begin{figure}[t]
\centering
\begin{tabular}{cc}
\includegraphics[width=0.43\linewidth,keepaspectratio=true,angle=0]{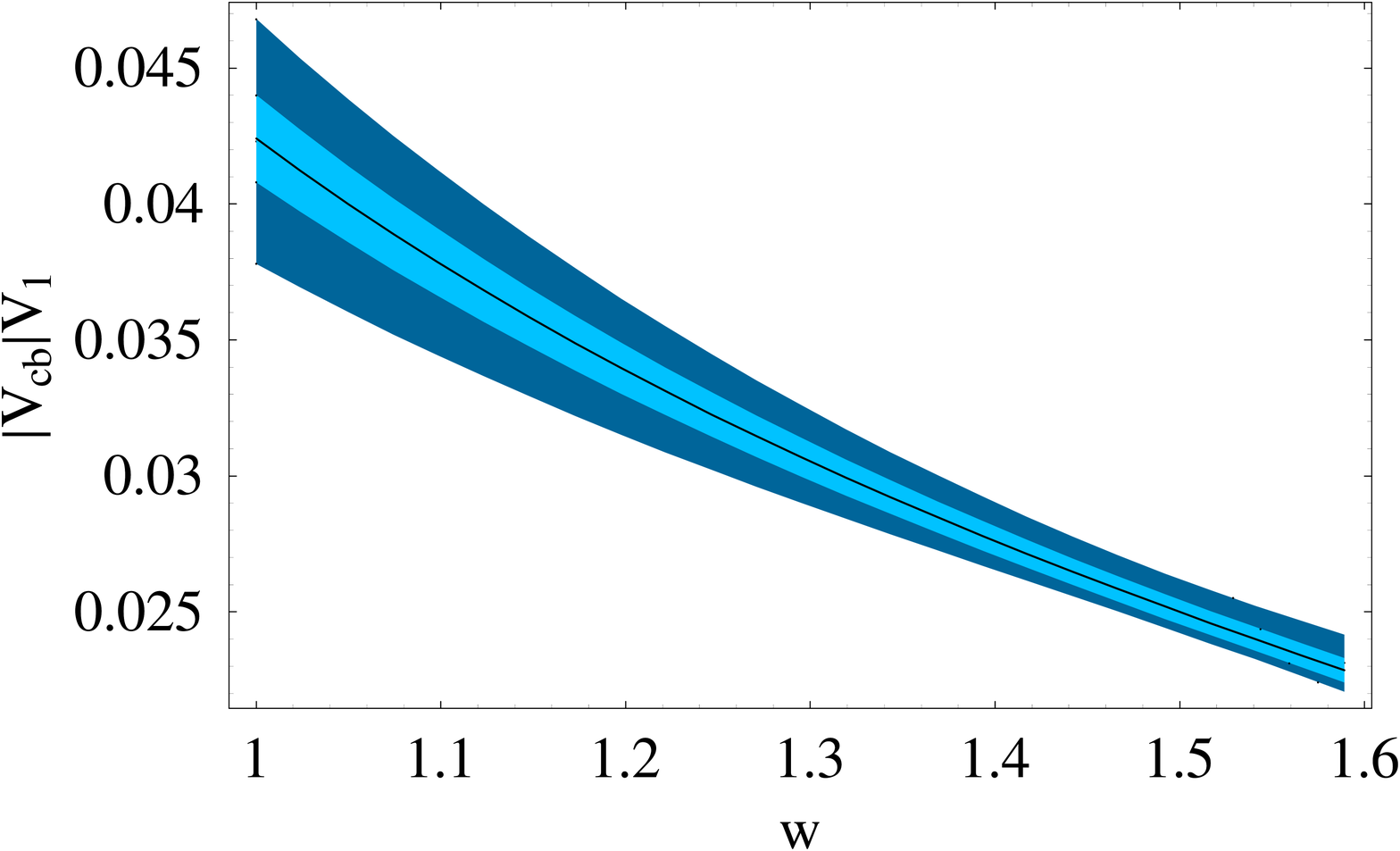} &
\includegraphics[width=0.43\linewidth,keepaspectratio=true,angle=0]{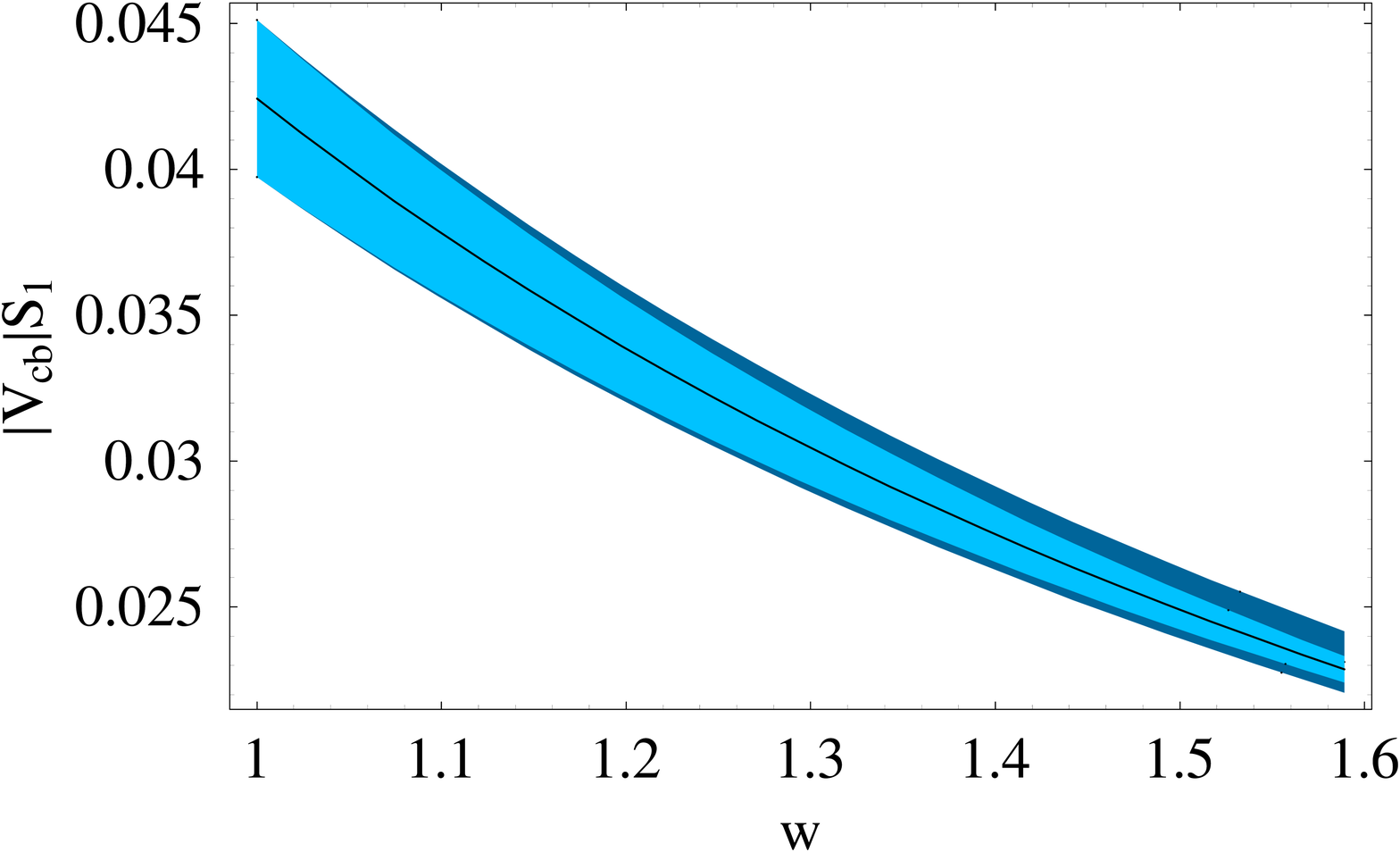} 
\end{tabular}
\caption{Vector (left) and scalar (right) form factors corresponding to the $|V_{cb}|\mathcal{G}(1)$ and $\rho^2$
determinations of HFAG before (dark gray/dark blue)\cite{HFAG07} and after (gray/blue)\cite{HFAG08} ICHEP08.
With the new determination, the errors on $|V_{cb}|V_1$ and $|V_{cb}|S_1$ are smaller than $4\%$ and $7\%$, respectively.}
\label{Fig2}
\end{figure}

This parametrization has been used in Ref.\cite{CLN98}
with the choice $t_0=q^2_\textrm{max}=(M_{B}-M_{D})^{2}$,
together with heavy-quark spin symmetry inputs,
to derive the following ansatz for the vector form factor:
\begin{align}
f_{+}(q^2)\equiv\frac{M_{B}+M_{D}}{2\sqrt{M_{D}M_{B}}}V_{1}(q^2),\quad
V_{1}(q^2)=\mathcal{G}(1)\left[  1-8\rho^{2}z(q^2,t_0)+(51\rho^{2}-10)z(q^2,t_0)^{2}-(252\rho^{2}-84)z(q^2,t_0)^{3}\right],
\label{Eq4}
\end{align}
where $V_1$ is defined such that it reduces to the Isgur-Wise function in the heavy-quark limit
and $\mathcal{G}(1)\equiv V_{1}(q^2_\textrm{max})$.
The parameters $|V_{cb}|\mathcal{G}(1)$ and $\rho^2$ can be determined from $\BDLN$ experimental data.
Before this summer, the HFAG averages \cite{HFAG07} based on BELLE, CLEO, and ALEPH data read:
$|V_{cb}|\mathcal{G}(1)=(42.3\pm4.5)\times10^{-3}$ and $\rho^2=1.17\pm0.18$ (with a $|V_{cb}|\mathcal{G}(1)$-$\rho^2$ correlation of $0.93$).
The recent BABAR results \cite{BABAR08_BDLNuT} and \cite{BABAR08_BDLNuU} have now been included,
leading to a substantial improvement \cite{HFAG08}:
$|V_{cb}|\mathcal{G}(1)=(42.4\pm0.7\pm1.4)\times10^{-3}$ and $\rho^2=1.19\pm0.04\pm0.04$ (with $|V_{cb}|\mathcal{G}(1)$-$\rho^2$ correlation $0.88$).
The old and new vector form factors are compared in Fig.\ref{Fig2} (left),
where we have defined as usual $w=(M_{B}^{2}+M_{D}^{2}-q^2)/(2M_{B}M_{D})$.

For the scalar form factor, we adopt the ansatz of Ref.\cite{Hill06}:
\begin{align}
f_{0}(q^2)\equiv\frac{(w+1)\sqrt{M_{D}M_{B}}}{M_{B}+M_{D}}S_{1}(q^2)
=\frac{1}{z(q^2,M_{1}^{2})z(q^2,M_{2}^{2})\phi_{0}(q^2,t_0)}\left[  a_{0}^{0}(t_0)+a_{1}^{0}(t_0)\,z(q^2,t_0)\right],
\label{Eq5}
\end{align}
where $t_0=(M_B+M_D)^2\left( 1-\sqrt{1-(M_B-M_D)^2/(M_B+M_D)^2}\,\right)$
such that $|z|_\textrm{max}$ is minimized,
$M_{1}=6.700\gev$ and $M_{2}=7.108\gev$ \cite{EichtenQ94} are the subthreshold poles,
and $\phi_0$ is obtained from Eq.(10) of Ref.\cite{Hill06} setting $Q^{2}=0$~and~$\eta=2$:
\begin{equation}
\phi_0(q^2,t_0)=\sqrt{\frac{2(M_{B}^2-M_{D}^2)^{2}}{16\pi}}
\frac{\sqrt{(M_{B}+M_{D})^{2}-q^2}}{((M_{B}+M_{D})^{2}-t_0)^{1/4}}
\frac{z(q^2,0)^2}{(q^2)^2}
\left(\frac{z(q^2,t_0)}{t_0-q^2}\right)^{-1/2}
\left(\frac{z(q^2,(M_{B}-M_{D})^{2})}{(M_{B}-M_{D})^{2}-q^2}\right)^{-1/4}.
\end{equation}
Following \cite{NTW08}, we truncate the series (\ref{Eq3}) after the first two terms.
This is motivated by the fact that $|z|_\textrm{max}=0.032$
and that a similar parametrization for $f_+$, when fitted to experimental data,
produces the same result as Eq.(\ref{Eq4}) in very good approximation \cite{NTW08}.
Then, $|V_{cb}|a^0_0(t_0)$ and $|V_{cb}|a^0_1(t_0)$ are determined imposing the conditions
(i) $|V_{cb}|S_1(0)=|V_{cb}|V_1(0)$ and (ii) $|V_{cb}|S_1(q^2_\textrm{max})=(4.24\pm0.27)\%$
(corresponding to $|V_{cb}|=(41.6\pm0.6)\times10^{-3}$ from $B\to X_c\ell\nu$ \cite{PDG08}
and $S_{1}(q^2_\textrm{max})=1.02\pm0.05$ from HQET \cite{NTW08}).
The scalar form factors obtained in this way from the old and new $|V_{cb}|V_1$ are not very different,
as one can see on Fig.\ref{Fig2} (right).

\begin{figure}[t]
\centering
\begin{tabular}{cc}
\includegraphics[width=0.43\linewidth,keepaspectratio=true,angle=0]{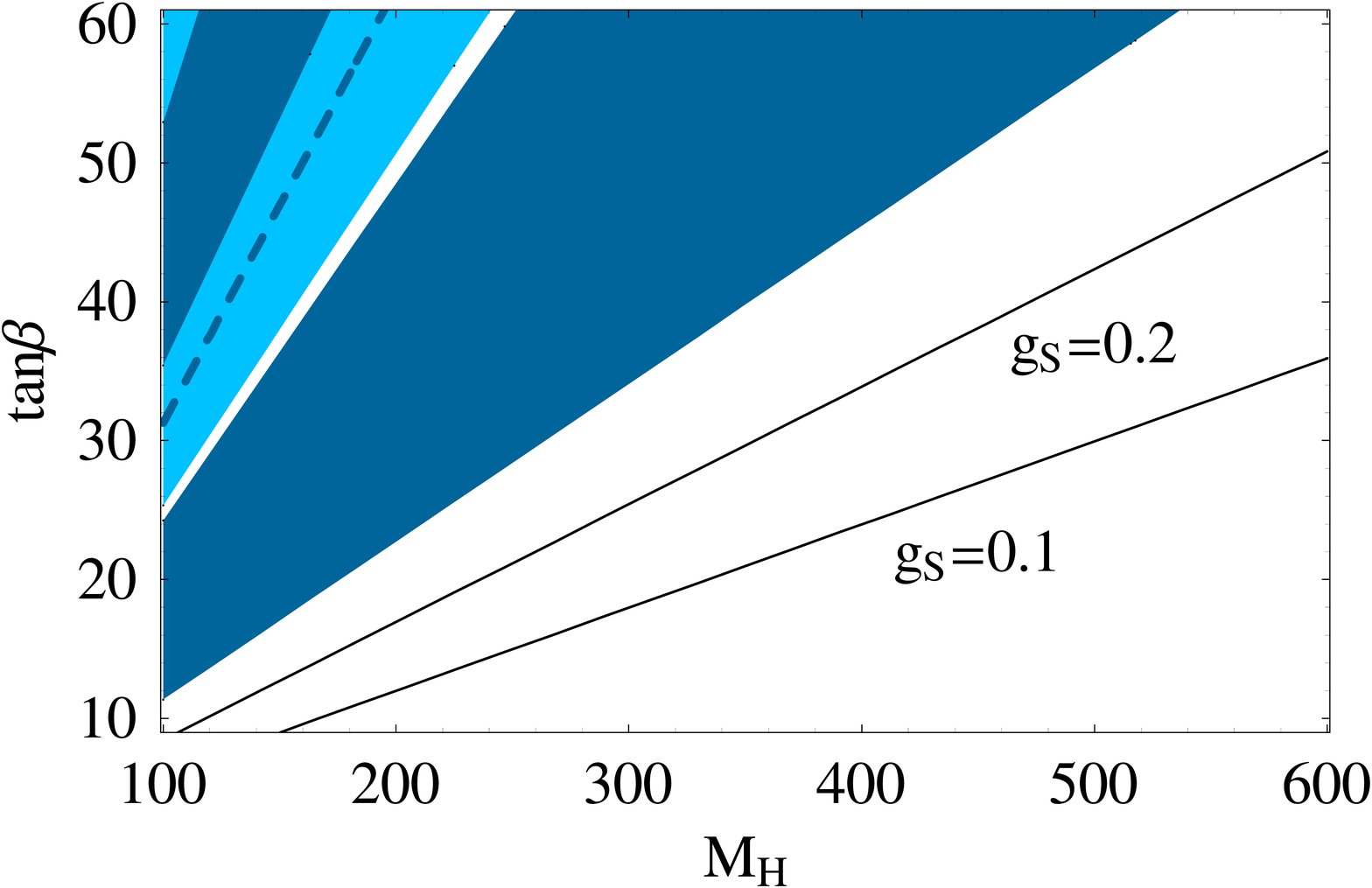} &
\includegraphics[width=0.43\linewidth,keepaspectratio=true,angle=0]{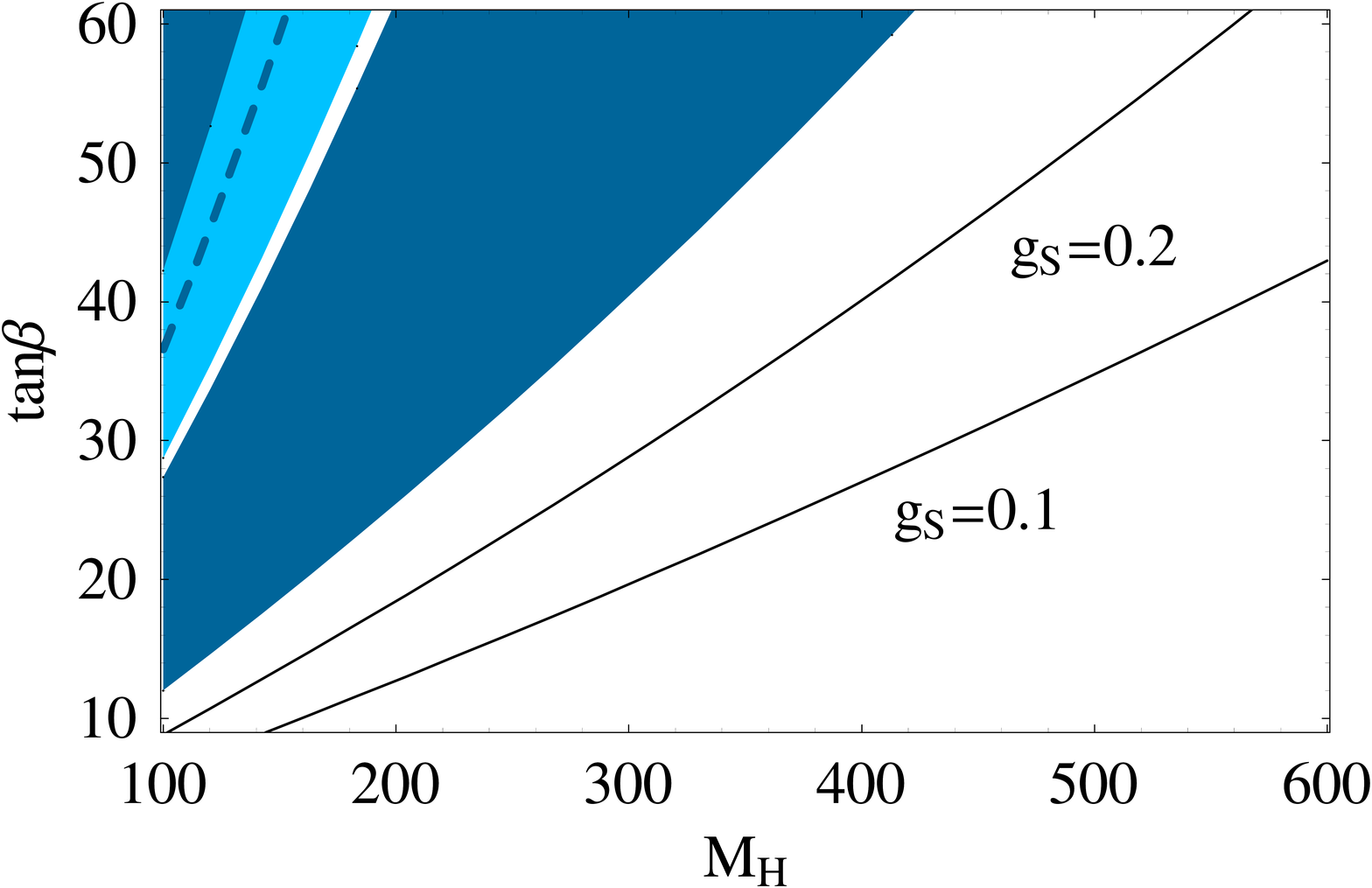} 
\end{tabular}
\caption{$95\%$ C.L. exclusion zones in the $(M_H,\tan\beta)$ plane from
$\mathcal{B}(B\rightarrow\tau\nu)$ (dark gray/dark blue) and $R$ (gray/blue)
in a 2HDM (left) and in the MSSM with $\varepsilon_{0}=0.01$ and $\varepsilon_{\tau}\simeq0$ (right).
The exclusion limits are directly read from the gray (blue) bands in Fig.\ref{Fig1}.
They differ from those usually found in the literature in that
experimental and theory errors are not simply added in quadrature
and the dependence of the errors on the $H^+$ contribution is taken into account.}
\label{Fig3}
\end{figure}

The recent progress on $|V_{cb}|V_1$ allows to reduce the errors on the SM predictions
for the two $\BDTN$ branching fractions:
$\mathcal{B}(B^-\rightarrow D^0\tau^-\bar\nu)^{SM}=\left(0.70^{+0.06}_{-0.05}\right)\%$,
$\mathcal{B}(\bar B^0\rightarrow D^+\tau^-\bar\nu)^{SM}=\left(0.65^{+0.06}_{-0.05}\right)\%$
(differing essentially due to $\tau_{B^0}\not=\tau_{B^+}$).
The errors from $|V_{cb}|V_1$, however, already cancel to a large extent in the ratio $R\equiv\mathcal{B}(\BDTN)/\mathcal{B}(\BDLN)$,
which is why the nice improvement in Fig.\ref{Fig2} has little impact on Fig.\ref{Fig1} (right),
already dominated by the error on $S_{1}(q^2_\textrm{max})$: $R^{SM}=0.31\pm0.02$.
This estimation is compatible with the one obtained from lattice methods: $R^{SM}_{latt}=0.28\pm0.02$ \cite{KamenikM08}.
Note that replacing condition (ii) by a constraint on $S_1(q^2_\textrm{max})/V_1(q^2_\textrm{max})$ from HQET
would lead to a similar error on $R$.
Still, an interesting $95\%$ C.L. bound on $g_S$ can already be obtained from $R$: $g_S<1.79$,
complementary to the bounds from $\mathcal{B}(B\rightarrow\tau\nu)$: $g_S<0.36\ \cup\ 1.64<g_S<2.73$.
The corresponding exclusion zones in the $(M_{H},\tan\beta)$ plane are depicted in Fig.\ref{Fig3}.
The error assigned to $S_{1}(q^2_\textrm{max})$ is quite conservative, so the above constraints are robust.
At the three-sigma level, it is not possible to extract any interesting bound from $R$ yet,
but its experimental knowledge is expected to improve in the near future.
Its role to constrain $H^+$ effects will then of course  
depend on the new central value.
For the moment, a $15\%$ measurement with the same central value would exclude $g_S>0.29$ at the $95\%$~C.L..

\section{\boldmath$B\rightarrow D\tau\nu$ DIFFERENTIAL
DISTRIBUTIONS\unboldmath}

If a hint for a charged Higgs boson is seen at the branching fraction level, $\BDTN$ has a great advantage
over $B\to\tau\nu$: it allows to analyze the same data points on a differential basis,
better suited to discriminate between effective scalar-type interactions and other effects.
The $d\Gamma(\BDTN)/dq^2$ distribution, in particular, has already been studied in great detail \cite{distr}.
The polarization of the $\tau$ is also known as a $H^{+}$ analyzer \cite{pol},
yet it requires the knowledge of the $\tau$ momentum, which cannot be accessed at $B$ factories
as the $\tau$ does not travel far enough for a
displaced vertex and decays into at least one more neutrino.

\begin{figure}[t]
\centering
\begin{tabular}{cc}
\includegraphics[width=0.43\linewidth,keepaspectratio=true,angle=0]{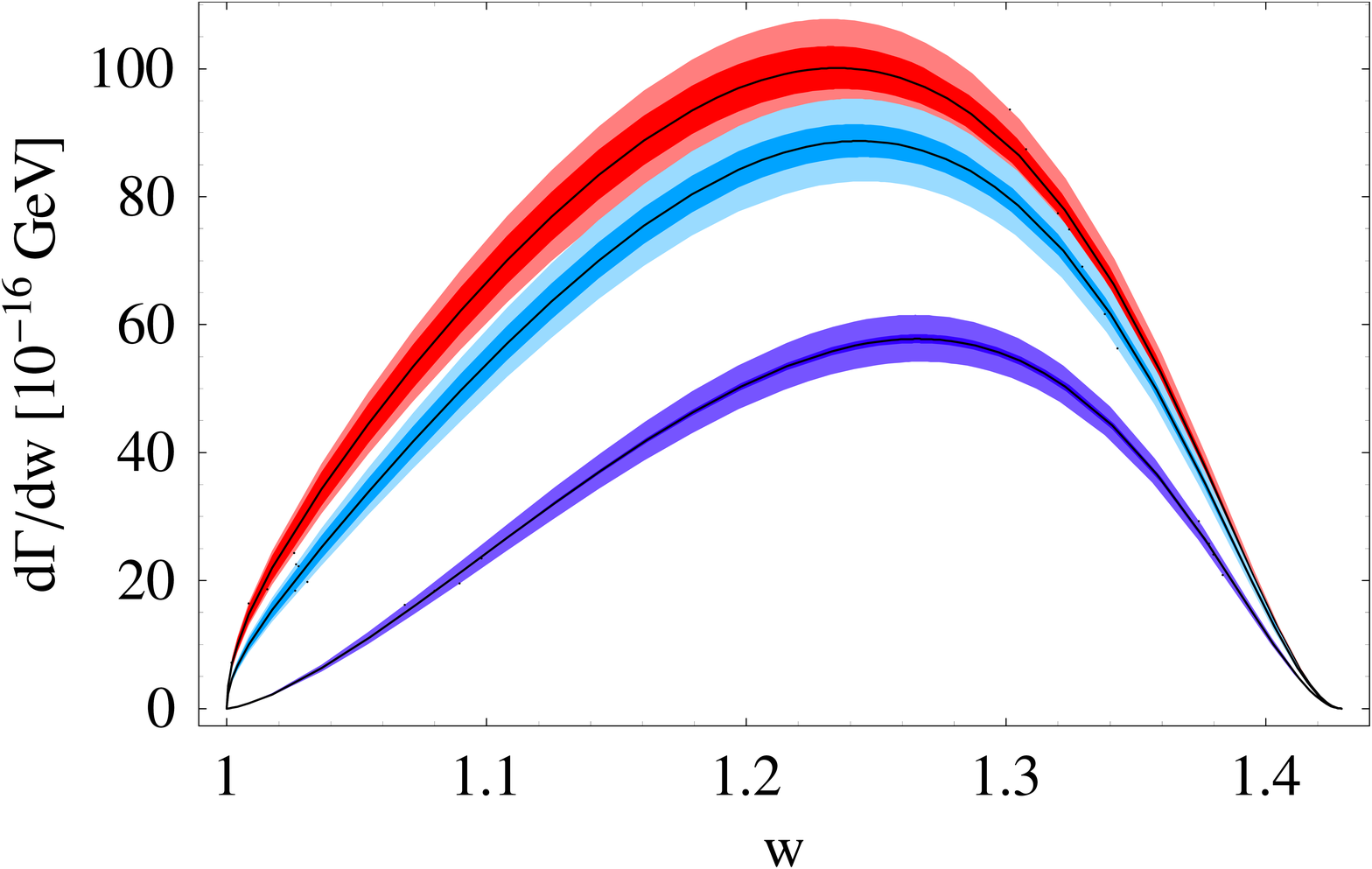} &
\includegraphics[width=0.43\linewidth,keepaspectratio=true,angle=0]{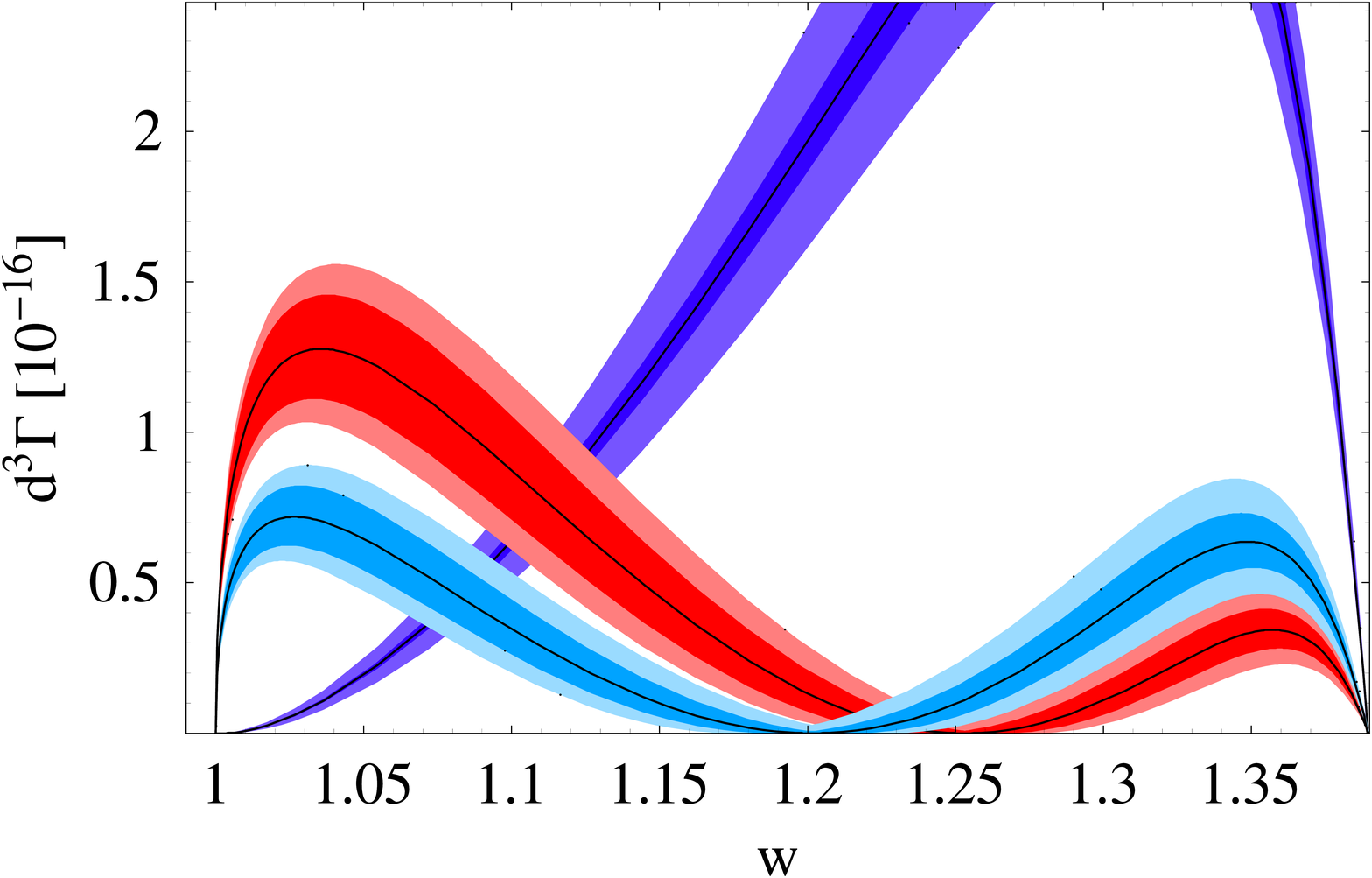} 
\end{tabular}
\caption{$d\Gamma(\BDTN)/dw$ (left) and $d\Gamma(B\to D\nu\tau[\to\pi\nu])/dE_\pi d\cos\theta_{D\pi} dw$
with $E_\pi=1.8\gev$ and $\cos\theta_{D\pi}=-1$ (right) for $g_S=0$ (gray/red), $g_S=0.35$ (light gray/light blue), and $g_S=1.75$ (dark gray/dark blue).
These values are still allowed by $\mathcal{B}(B\rightarrow D\tau\nu)$ and $\mathcal{B}(B\rightarrow\tau\nu)$ at the $95\%$ C.L..
The various curves have been obtained using the more recent HFAG vector form factor \cite{HFAG08}.
The lighter bands take all errors into account, while the darker bands only take into account
the error on $S_1(q^2_\textrm{max})$.
One could of course also normalize the above differential distributions to $d\Gamma(\BDLN)/dw$
to reduce the impact of the errors on $f_+$.}
\label{Fig4}
\end{figure}

A straightforward way to nevertheless exploit the sensitivity of the $\tau$ polarization
to $H^+$ effects and at the same time retain the information from the $q^2$ spectrum
is to look at the subsequent decay of the $\tau$ into a pion and a neutrino \cite{NTW08}.
The direction of the pion is indeed directly correlated with the polarization of the $\tau$.
Integrating over the neutrino momenta, we end up with a triple differential decay distribution
$d\Gamma(B\to D\nu\tau[\to\pi\nu])/dq^2 dE_\pi d\cos\theta_{D\pi}$.
An explicit formula is given in Eqs.(9-11) of Ref.\cite{NTW08} (with $F_V\equiv f_+$ and $F_S\equiv f_0$).
Its sensitivity to $g_S$ is illustrated in Fig.\ref{Fig4} for $E_\pi=1.8\gev$ and $\cos\theta_{D\pi}=-1$.
For comparison, we also display the $q^2$ spectra corresponding to the same $g_S$ values.
Of course, in practice, one should not fix $E_\pi$ or $\theta_{D\pi}$,
but rather perform a (unbinned) maximum likelihood fit
of the triple differential decay distribution to the available data points.
The information from the $q^2$ spectrum in the dominant $\tau\to\ell\nu\bar\nu$
decay channel should also be included in the fit to make the most out of experimental data.

\section{CONCLUSION}

The form factors $f_+(q^2)$ and $f_0(q^2)$ in the $B\rightarrow D\tau\nu$ transition are under good control.
As a result, the ratio $R\equiv\mathcal{B}(\BDTN)/\mathcal{B}(\BDLN)$ can be predicted
with $7\%$ accuracy in the SM: $R^{SM}=0.31\pm0.02$,
where the $5\%$ uncertainty on the scalar form factor at zero recoil
$S_1(q^2_\textrm{max})$ is the main error source.
This allows to derive useful constraints on the effective $H^+$ coupling $g_S$.
Together with the constraints from $\mathcal{B}(B\rightarrow\tau\nu)$, we obtain:
$g_S<0.36\ \cup\ 1.64<g_S<1.79$, i.e., the window around $g_S=2$
left over by $\mathcal{B}(B\rightarrow\tau\nu)$ is now nearly completely excluded by $R$ alone.
These bounds should be strengthened soon thanks to the current considerable experimental efforts on both modes.
In this respect, one should pay particular attention to the $B\rightarrow D\tau\nu$ differential distributions
as these are especially well-suited to discriminate between effective scalar interactions
and other types of effects and,
if the former are seen, to extract the coupling $g_S$ with good precision.

\begin{acknowledgments}

It's a pleasure to thank my collaborators
Ulrich Nierste and Susanne Westhoff. Discussions with Matthias Steinhauser about the ratio $m_c/m_b$ and with Christoph Schwanda and Laurenz Widhalm
about experimental issues are also warmly acknowledged. Work
supported by the DFG grant No. NI 1105/1-1, by the DFG-SFB/TR9, and by the EU
contract No. MRTN-CT-2006-035482 (FLAVIAnet).

\end{acknowledgments}


\begin{thebibliography}{99}

\bibitem{Hou93} W.~Hou, Phys. Rev. \textbf{D48}, 2342 (1993).

\bibitem{AkeroydR03} M.~S.~Carena, D.~Garcia, U.~Nierste and C.~E.~M.~Wagner, Nucl.\ Phys. {\bf B577}, 88 (2000);
A.~J.~Buras, P.~H.~Chankowski, J.~Rosiek and L.~Slawianowska, Nucl.\ Phys. {\bf B659}, 3 (2003);
A.~Akeroyd and S.~Recksiegel, J. Phys. G \textbf{29}, 2311 (2003);
H.~Itoh, S.~Komine and Y.~Okada, Prog.\ Theor.\ Phys.\  {\bf 114}, 179 (2005).

\bibitem{Pheno}
For a recent global phenomenological analysis, see for example D.~Eriksson, F.~Mahmoudi and O.~Stal, arXiv:0808.3551 [hep-ph].

\bibitem{PDG08} C.~Amsler \emph{et al.} (Particle Data Group), Phys. Lett. \textbf{B667}, 1 (2008).

\bibitem{HFAG08} Heavy Flavor Averaging Group, update after ICHEP08, http://www.slac.stanford.edu/xorg/hfag/.

\bibitem{HPQCD05} A.~Gray \emph{et al.} (HPQCD Collaboration), Phys. Rev. Lett. \textbf{95}, 212001 (2005).

\bibitem{BABAR08_BDTauNu} B.~Aubert \emph{et al.} (BABAR Collaboration), Phys.\ Rev.\ Lett. \textbf{100}, 021801 (2008).

\bibitem{HFAG07} Heavy Flavor Averaging Group, arXiv:0808.1297 [hep-ex].

\bibitem{mcmb} K.~Chetyrkin, J.~K\"{u}hn and M.~Steinhauser,
Comput.\ Phys.\ Commun. \textbf{133}, 43 (2000). Input values: J.~K\"{u}hn,
M.~Steinhauser and C.~Sturm, Nucl. Phys. \textbf{B778}, 192 (2007).

\bibitem{zpara} See for example the discussion by F.~J.~Yndurain, arXiv:hep-ph/0212282.

\bibitem{CLN98} I.~Caprini, L.~Lellouch and M.~Neubert, Nucl. Phys. \textbf{B530}, 153 (1998).
See also the earlier related works:
C.~Bourrely, B.~Machet and E.~de Rafael, Nucl. Phys. {\bf B189}, 157 (1981);
C.G.~Boyd, B.~Grinstein and R.F.~Lebed, Phys. Rev. \textbf{D56}, 6895 (1997).

\bibitem{BABAR08_BDLNuT} B.~Aubert \emph{et al.} (BABAR Collaboration), these proceedings, arXiv:0807.4978 [hep-ex].
\bibitem{BABAR08_BDLNuU} B.~Aubert \emph{et al.} (BABAR Collaboration), arXiv:0809.0828 [hep-ex].

\bibitem{Hill06} R.~Hill, arXiv:hep-ph/0606023.

\bibitem{EichtenQ94} E.~Eichten and C.~Quigg, Phys. Rev. \textbf{D49}, 5845 (1994).

\bibitem{NTW08}U.~Nierste, S.~Trine and S.~Westhoff, Phys. Rev.
\textbf{D78}, 015006 (2008), arXiv:0801.4938 [hep-ph].

\bibitem{KamenikM08} J.F.~Kamenik and F.~Mescia, Phys. Rev. \textbf{D78}, 014003 (2008), arXiv:0802.3790[hep-ph].

\bibitem{distr} B.~Grzadkowski and W.~Hou, Phys. Lett. \textbf{B283}, 427 (1992);
K.~Kiers and A.~Soni, Phys. Rev. \textbf{D56}, 5786 (1997).

\bibitem{pol} J.~Kalinowski, Phys. Lett. \textbf{B245}, 201 (1990);
D.~Atwood, G.~Eilam and A.~Soni, Phys.\ Rev.\ Lett.\  {\bf 71}, 492 (1993);
Y.~Grossman and Z.~Ligeti,Phys.\ Lett.\ {\bf B332}, 373 (1994);
M.~Tanaka, Z. Phys. \textbf{C67}, 321 (1995);
R.~Garisto, Phys.\ Rev.\ {\bf D51}, 1107 (1995);
G.~H.~Wu, K.~Kiers and J.~N.~Ng, Phys.\ Rev. {\bf D56}, 5413 (1997).

\end{thebibliography}
\end{document}